\documentclass[aps,prd,12pt,groupedaddress,preprintnumbers,floatfix,nofootinbib]{revtex4}

\usepackage{amsmath}
\usepackage{graphicx}
\usepackage{color}
\usepackage{dcolumn}% Align table columns on decimal point
\usepackage{bm}% bold math

\begin{document}

\preprint{DCP-09-04}

\title{Electroweak scale neutrinos and decaying dark matter}

\author{
Alfredo Aranda,$^{1,2}$\footnote{Electronic address:fefo@ucol.mx} 
and Francisco J. de Anda$^{3}$} 

\affiliation{$^1$Facultad de Ciencias - CUICBAS,\\ 
  Universidad de Colima, M\'exico\\
  $^2$Dual C-P Institute of High Energy Physics, M\'exico \\
  $^3$Departamento de F\'isica - CUCEI, Universidad de Guadalajara, M\'exico}

\date{\today}

\begin{abstract}
We explore the scalar phenomenology of a model of electroweak scale neutrinos that incorporates
the presence of a lepton number violating singlet scalar. An analysis of
the pseudoscalar-Majoron  field associated to this singlet field is carried out in order to verify 
the viability of the model and to restrict its parameter space. In particular we study the 
Majoron decay $J \to \nu \nu$ and use the bounds on the Majoron mass and width obtained in a modified Majoron Decaying Dark Matter scenario. 
\end{abstract}

\maketitle
%\section{Introduction}
It would be very exciting if the mechanism associated to neutrino mass generation could be testable at
collider energies. This has recently motivated efforts into the creation and study of models that invoke the 
existence of electroweak scale right-handed neutrinos~\cite{Graesser:2007pc,Hung:2006ap,Kersten:2007vk,de Gouvea:2007uz,Aranda:2007dq,Bu:2008fx,Hung:2008ga,delAguila:2008hw,Aranda:2008ab,Aranda:2008ed,Perez:2009mu}, where the seesaw mechanism
in some of its several realizations is then used to create small neutrino masses~\cite{Schechter:1980gr}. 
The purpose of this letter is to explore the
viability of a model~\cite{Aranda:2007dq} that incorporates a minimum set of additions to the Standard Model (SM), 
namely the addition
of three right-handed neutrinos and a lepton number violating singlet scalar, through the implementation of cosmological constraints for Majorons~\cite{Cline:2009sn}.
 
The letter is organized as follows: first we present the salient features of the model described in~\cite{Aranda:2007dq}. 
We then describe the analysis performed using the cosmological constraints on the Majoron mass and width 
in order to constrain the parameter space of the model, and thus check its viability. Lastly we conclude with 
some final remarks. 

The model adds to the SM particle content the following: three right-handed neutrinos ($N_{Ri}$, $i=1,2,3$) 
with a mass of electroweak (EW) scale size, and a SM singlet
complex scalar field ($\eta$) that breaks global U(1)$_L$ symmetry. The field $\eta$ has lepton number $-2$. The 
relevant terms in the lagrangian are given by
\begin{eqnarray}\label{lagrangian} 
\mathcal{L}_{\nu H} &=& \mathcal{L}_{\nu y}-V \ , \\ \label{yukawa}
\label{lnuy} \mathcal{L}_{\nu y} &=& -y_{\alpha i}\overline{L}_\alpha \tilde{\Phi}N_{Ri}-\frac{1}{2}Z_{ij}
\eta \overline{N}^c_{Ri}N_{Rj}+ h.c. \ , \\ \label{potential}
V &=& \label{sca} \mu_D^2\Phi^\dagger\Phi+\frac{\lambda}{2}(\Phi^\dagger\Phi)^2+\mu_S^2\eta^*\eta+
\lambda'(\eta^*\eta)^2\\ \nonumber &+& \kappa(\eta\Phi^\dagger\Phi+\eta^*\Phi^\dagger\Phi)+
\lambda_m(\Phi^\dagger\Phi)(\eta^*\eta) \ ,
\end{eqnarray}
where $\tilde{\Phi}=i\sigma^2\Phi^*$, $(\Psi_R)^c=P_L \Psi^c$,
$\Psi^c=C\gamma^0\Psi^*$, and where
\begin{eqnarray}\label{scalars}
\Phi= \left(\begin{array}{c}0 \\ \frac{\phi^0+v}{\sqrt{2}}\end{array}\right),\ 
\eta=\frac{\rho+u+i J}{\sqrt{2}} \ ,
\end{eqnarray}
with $v (u)$ denoting the vacuum expectation value (vev) of $\Phi (\eta)$.
Some remarks: the model assumes that all mass scales come from spontaneous symmetry breaking (SSB) and that
there is only one high energy scale, namely the EW scale. Thus both $v$ and $u$ are of this order. The
potential includes a term that explicitly breaks U(1)$_L$. We work under the assumption that this explicit
breaking is very small, i.e. $|\kappa|\ll $~EW scale. Note that since the Majoron mass in the model is given by
\begin{eqnarray}\label{majoronmass}
M_J^2 = \frac{-\kappa v^2}{\sqrt{2}u} \ ,
\end{eqnarray}
the cosmological conditions will constrain $\kappa$. As we show below, the assumptions of the model satisfy such
constraint.

In~\cite{Aranda:2007dq} neutrino masses were obtained using the seesaw mechanism. The main observation is that
since the right-handed neutrinos are of EW scale size, it is required for the Yukawa couplings $y_{\alpha i}$ in 
Eq.(\ref{yukawa}) to be small. In particular, considering the third family of SM fields and one
right-handed 
neutrino, the value obtained is $y_{\tau i} \le 10^{-6}$, i.e. of the same order as the electron Yukawa. Further explanation of the relative sizes of these (and all other) Yukawa couplings must be provided by other means (one example is the use of horizontal flavor symmetries). This issue is not addressed in this letter.

In all of the analysis presented in~\cite{Aranda:2007dq} it was required that the right-handed neutrino mass scale
be of EW scale size. Looking at Eq.(\ref{yukawa}) this implies that, since $u \sim $ O(EW), the unknown factors $Z_{ij}$ 
must be of O(1), but no direct constraint was obtained. If imposing additional constraints requires $Z_{ij}$ to be extremely small or large, the model would not be viable as it stands. 

We now explore whether or not this model can satisfy the cosmological constraints on the mass and width of the 
Majoron. To do so we use the results obtained from the Majoron decaying dark matter (DDM) idea~\cite{Berezinsky:1993fm}, within a
modified $\Lambda$CDM cosmological model in which a Majoron with a mass in the keV range is the dark matter particle~\cite{Lattanzi:2007ux}. Using the cosmological microwave observations from the Wilkinson Microwave Anisotropy Probe (WMAP)~\cite{Spergel:2006hy}, and
assuming that all the dark matter is composed of Majorons, the marginalized limits for $\Gamma_J$ and $M_J$
obtained in~\cite{Lattanzi:2007ux} are
\begin{eqnarray} \label{width}
  \Gamma_J < 1.3\times 10^{-19} \ \rm{sec}^{-1} \ , \\ \label{mass}
  0.12 \ \rm{keV} < M_J < 0.17 \ \rm{keV} \ .
\end{eqnarray}

Assuming for simplicity that $Z_{ij}=0$ for $i\neq j$, and defining $Z=Z_{11}\approx Z_{22}\approx Z_{33}$, i.e. assuming the right-handed neutrino to be family diagonal and
approximately proportional to the identity, we obtain the following expression for the Majoron decay to
neutrinos (for one family):
\begin{eqnarray}\label{decay}
  \Gamma_J = \frac{Z^2\sin^4\theta}{4\pi M_J} \left(1-\frac{4m_{\nu}^2}{M_J^2}\right)^{1/2}m_{\nu}^2 \ ,
\end{eqnarray}
where $\theta \approx 10^{-6}$ is the mixing angle in the neutrino sector (see~\cite{Aranda:2007dq}).

Saturating the bounds on the width and mass of the Majoron in Eqs.~(\ref{width}) and~(\ref{mass}) requires $Z$ 
to be of O($10^{-2}$). This together with the fact that $u \sim$ O(EW), which was taken to be anywhere 
from $10$~GeV to $1$~TeV, allows one to have EW size masses for the right-handed neutrinos and the model
fits naturally into this scenario. Furthermore, from Eq.~(\ref{majoronmass}) one immediately observes that $|\kappa|/(GeV) \sim 10^{-9} \ll 1$, as required by the model assumptions.

Note that in this model there is no contribution to the Majoron width from its possible decay to photons. 
Such contribution would come from an effective term of the form $g_{\gamma} J \epsilon^{\mu\nu\rho\sigma}F_{\mu\nu}F_{\rho\sigma}$, 
present in several models where a triplet is used in order to obtain neutrino masses. Thus the model satisfies the 
bounds presented in~\cite{Bazzocchi:2008fh} automatically.  

\begin{figure}[h]
  \includegraphics[width=12cm]{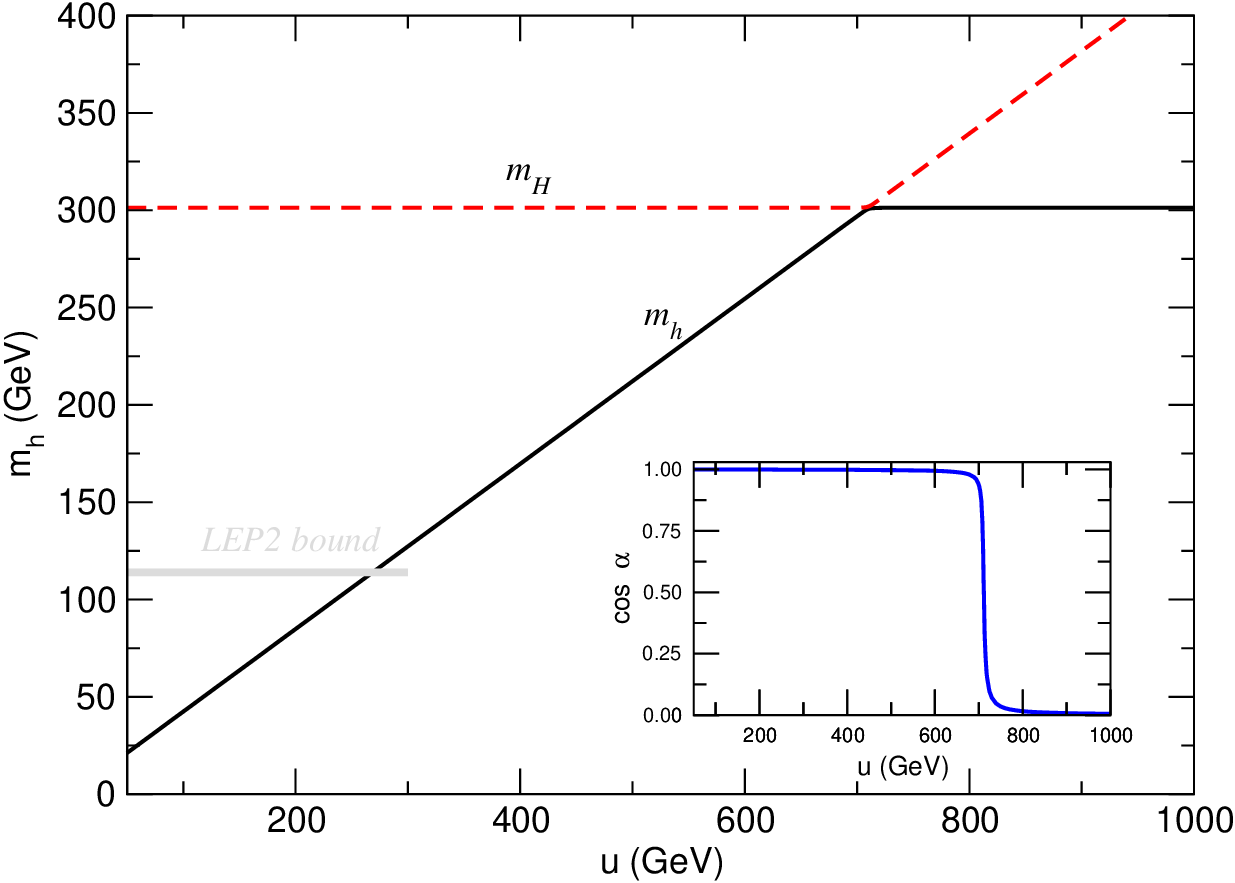}
  \caption{Scalar spectrum for the case described in the text.
The horizontal line in the lower left corresponds to the $114$~GeV lower bound from LEP2~\cite{Barate:2003sz}.}
  \label{figure1}
\end{figure}

\begin{figure}[ht]
  \includegraphics[width=8cm]{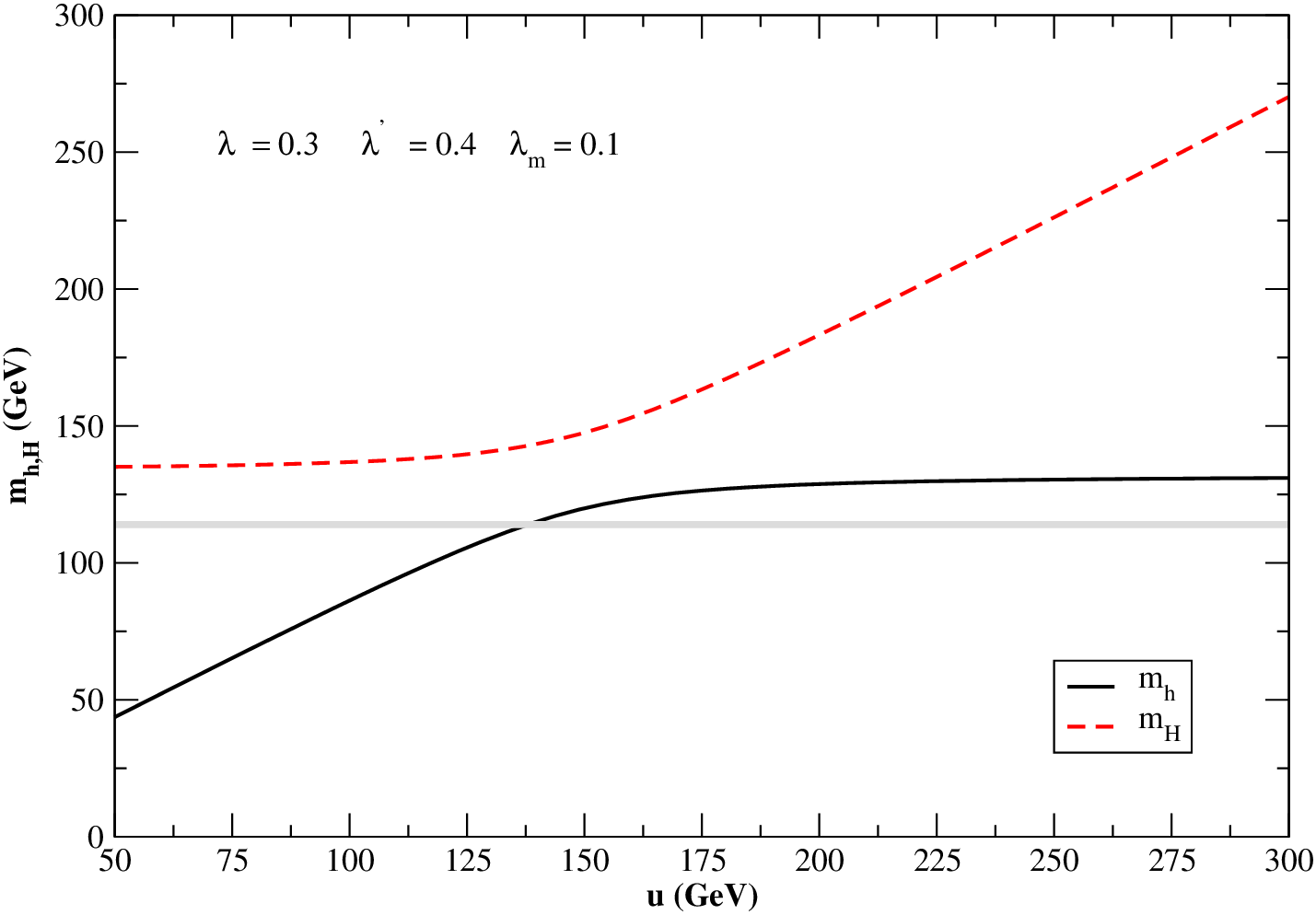}
  \includegraphics[width=8cm]{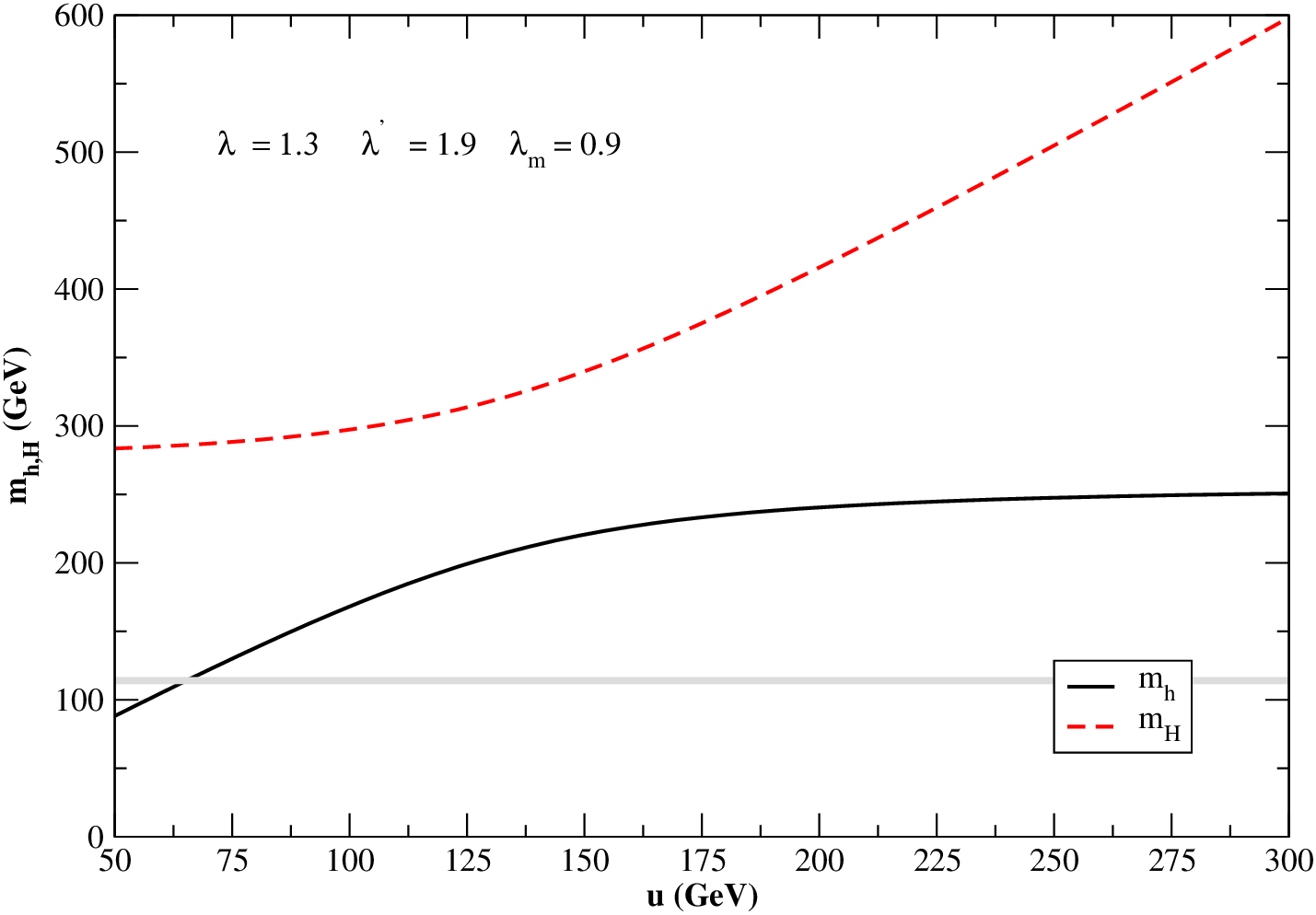}
  \\
  \includegraphics[width=8cm]{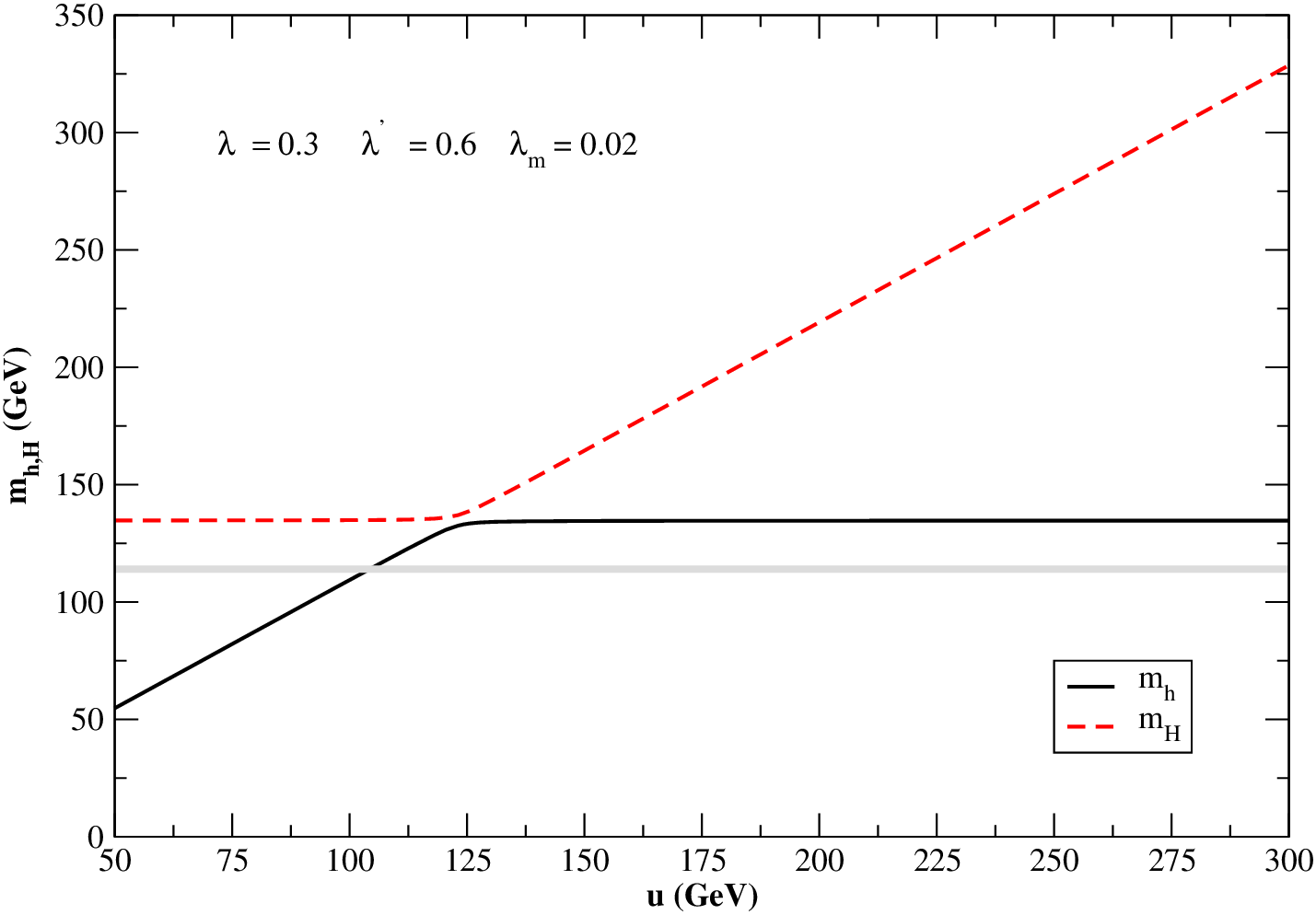}
  \includegraphics[width=8cm]{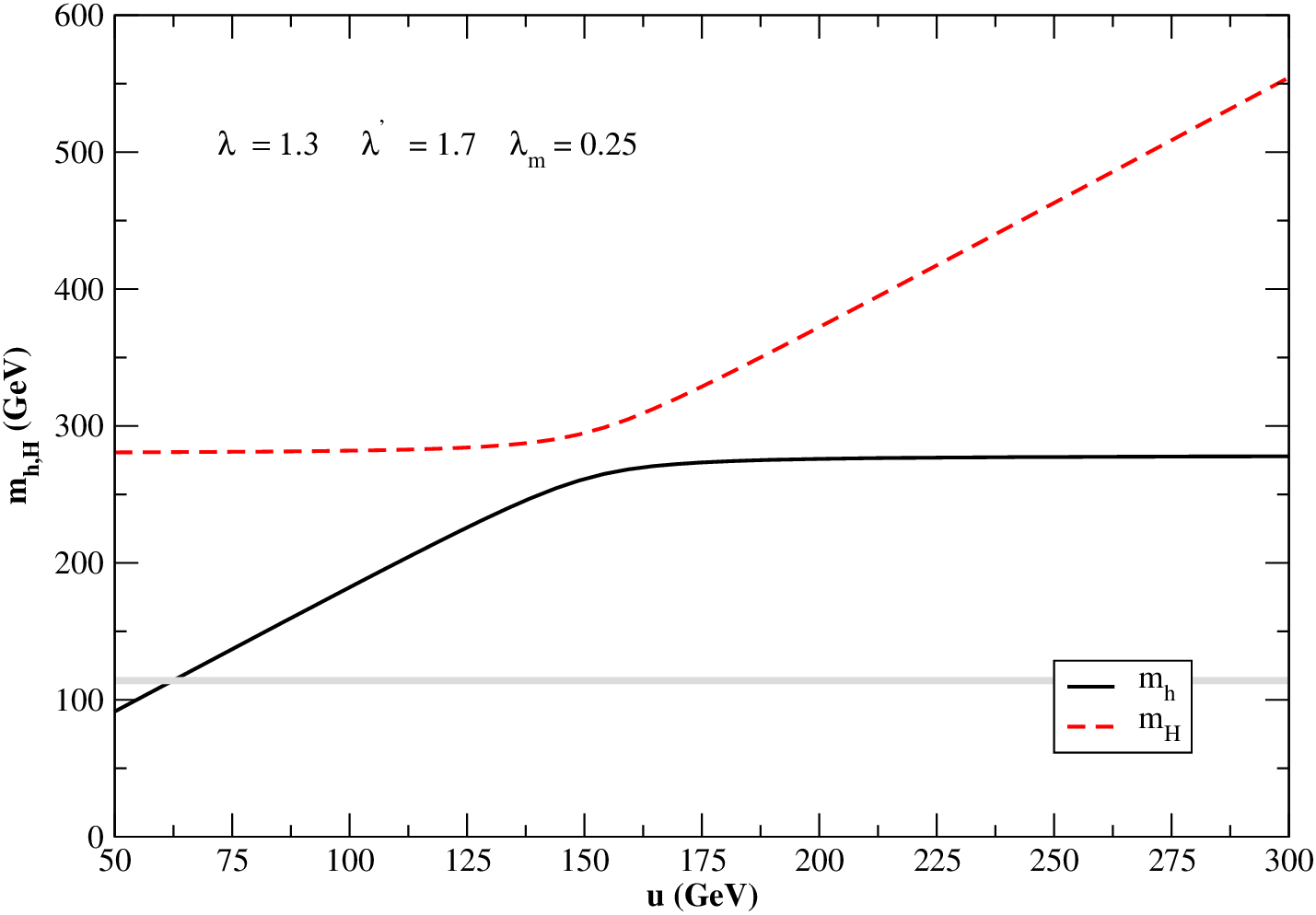}
  \caption{Scalar spectrum for four different choices of the parameters in the potential. In all four cases $\cos(\alpha)\sim 1$ only for $u \le 150$~GeV.
The horizontal line in corresponds to the $114$~GeV lower bound from LEP2~\cite{Barate:2003sz}.}
  \label{figure2}
\end{figure}

In order to study the viability of the scalar sector of the model one must explore its parameter space. We find
that it is possible to obtain sensible light Higgs masses for a wide range in the parameters, however, the mixing between the two scalar fields feeds into the Higgs production cross sections and has to be considered in determining whether or not the present scenario is of any relevance at a collider. Denoting the mixing angle by $\alpha$ (in the basis $(\phi^0 \ \rho)^T$) all SM processes get suppressed by a factor $\cos(\alpha)^{2}$~\cite{Aranda:2007dq}), thus we look for situations where $\cos(\alpha) \sim 1$ and show one such case in Figure~\ref{figure1}. The parameters used in this case are $\lambda = 1.5$, $\lambda^{\prime} = 0.09$, and $\lambda_m = 0.002$ which 
represent a typical set under these requirements, i.e. the quartic term associated to $\Phi$ must be larger than the one associated to $\eta$ and the mixing term must be smaller (no unexpected since in such a situation the doublet $\Phi$ dominates). Other choices of parameters also lead to acceptable light Higgs masses but with suppressed couplings for most of the $u$ range where we find that $\cos(\alpha) \sim 1$ for $u \le 150$~GeV (see Figure~\ref{figure2}).

\begin{figure}[ht]
  \includegraphics[width=12cm]{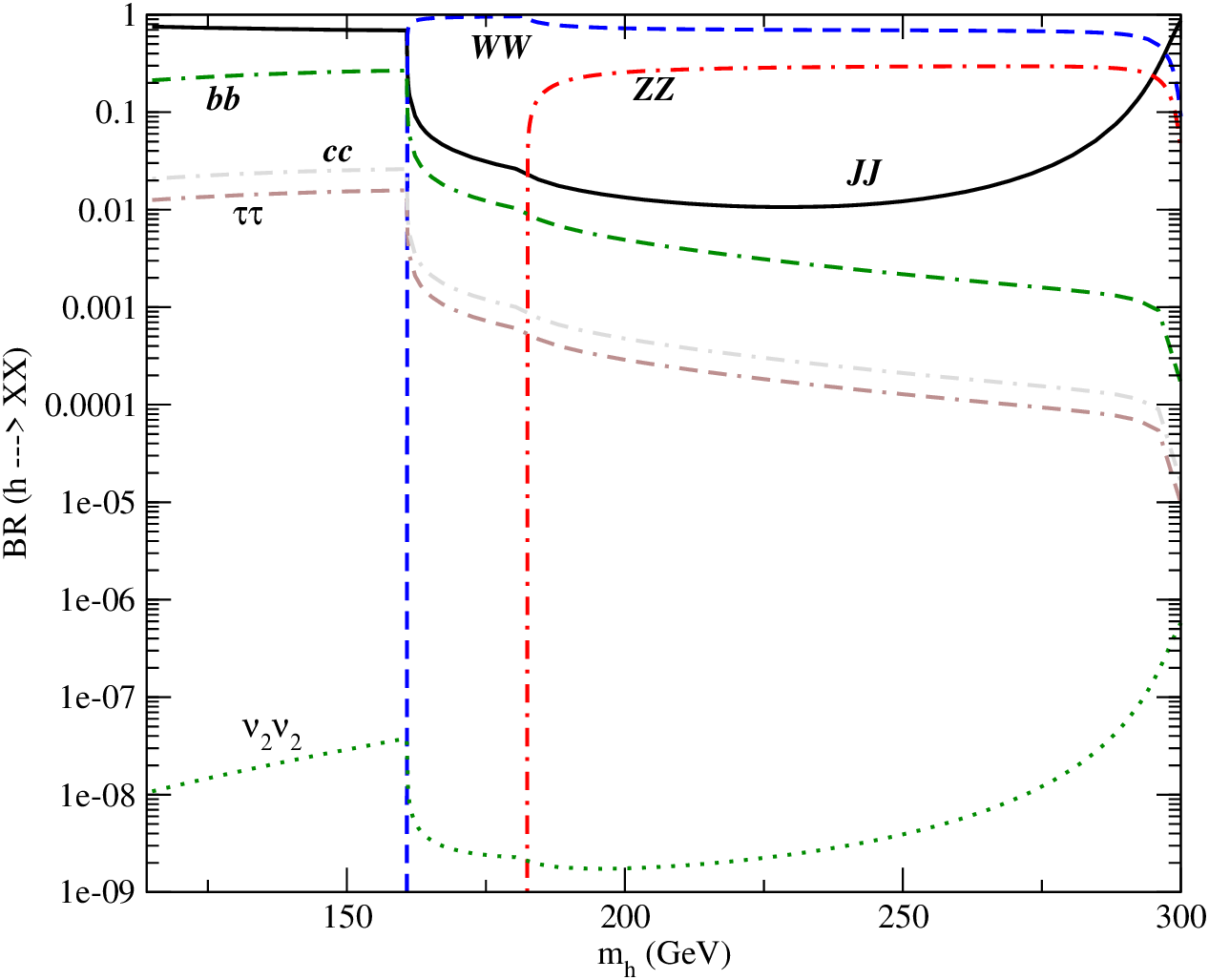}
  \caption{Branching ratios for the decay $h \to XX$ for the set of parameters
shown in Figure~\ref{figure1}.}
  \label{figure3}
\end{figure}

As discussed at the beginning of this letter, the main motivation for this model is to explore whether or not one can expect to probe the seesaw at colliders. Within the framework of this model the Higgs can decay, in addition to the SM channels, to a pair of Majorons $J$~\cite{Joshipura:1992hp} and to a pair of heavy neutrinos ($\nu_2$ in the notation of~\cite{Aranda:2007dq}). We have computed the branching ratio (BR) for each decay and find that the invisible Majoron decay dominates (BR $\sim 100\%$) throughout most of the parameter space.
Figure~\ref{figure3} shows the branching ratios for the set of parameters used in Figure~\ref{figure1} (we have plotted only the range of $m_h$ where $\cos(\alpha) \sim 1$). For the choices shown in Figure~\ref{figure2} the $h\to JJ$ channel has a BR$\sim 100\%$ for all the relevant range.

At the LHC the main production process, assuming only SM processes, is the gluon $-$ gluon fusion process. If the Higgs decays invisibly then, using this production process, one needs to look at its production in association with a jet ($h\to gg + {\rm jet}$). The signature will be that of the so-called mono-jet event and will be swamped by the pure QCD background. The vector-vector fusion process will also lead to a purely hadronic final signature with a huge background. Other production processes are the Higgs-strahlung and the associate production with top-quarks. In these two last cases the situation is better for a possible detection as they offer final leptonic states from W/Z decays~\cite{Godbole:2003it}. For example, the Higgs-strahlung with an associated Z production channel, can provide an observation of the invisible Higgs decay with over $5\sigma$ significance at the LHC for a Higgs with a mass in the range $114 - 140$~GeV and with a luminosity of only
$10$~fb$^{-1}$~\cite{Zhu:2005hv}. For larger Higgs masses it is also possible to discover the invisible Higgs up to a mass of $480$~GeV (at the $5\sigma$ level) if its invisible BR is $100 \%$~\cite{Eboli:2000ze}.

We conclude that the model is viable and well motivated. Its parameter space has been restricted by imposing constraints on the mass and width of the Majoron (assuming all dark matter is composed of Majorons) obtained from the cosmological microwave observations of WMAP. We verified the scalar spectrum of the model to be realistic once the constraints were imposed, and found that the scalar phenomenology is completely dominated by the invisible Higgs decay to two Majorons. We discussed the possible detection of such an invisible Higgs at the LHC and conclude that given the model parameters such discovery is possible.

\begin{acknowledgments}
The authors acknowledge support from CONACyT and SNI. 
\end{acknowledgments}

\end{document}